\ifx\pdfoutput\undefined 
\documentclass[aps,prl,showpacs,twocolumn,superscriptaddress]{revtex4}
\usepackage[dvips]{graphicx} 
\RequirePackage[dvipdfm,hyperindex]{hyperref}
\else 
\documentclass[aps,prl,showpacs,twocolumn,superscriptaddress]{revtex4}
\usepackage{graphicx}
\DeclareGraphicsExtensions{.pdf,.png,.jpg,.mps} 
\fi
\usepackage[english]{babel}
\usepackage{bm} 
\usepackage{color} 
\usepackage{amsmath} 
\usepackage{natbib}
\usepackage{SIunits} 
\usepackage{version}
\usepackage{longtable}

\newcommand{\cocp}{CoCp${}_2$}
 
\usepackage{epstopdf}

\begin{document}

\title{Graphene-mediated exchange coupling between cobaltocene and magnetic substrates}
\date{\today}

\author{S.~Marocchi}
\email[Electronic address: ]{simone.marocchi@unimore.it}
\affiliation{Dipartimento di Fisica, Universit\'a di Modena e Reggio Emilia, Via Campi 213/A, 41125 Modena, Italy.}
\affiliation{S3 - Istituto di Nanoscienze - CNR, Via Campi 213/A, 41125 Modena, Italy}
\author{P.~Ferriani}
\affiliation{Instit\"ut f\"ur Theoretische Physik und Astrophysik, Christian-Albrecht-Universit\"at zu Kiel, Leibnizstr. 15, 24098 Kiel, Germany}
\author{N.~M.~Caffrey}
\affiliation{Instit\"ut f\"ur Theoretische Physik und Astrophysik, Christian-Albrecht-Universit\"at zu Kiel, Leibnizstr. 15, 24098 Kiel, Germany}
\author{F.~Manghi}
\affiliation{Dipartimento di Fisica, Universit\'a di Modena e Reggio Emilia, Via Campi 213/A, 41125 Modena, Italy.}
\affiliation{S3 - Istituto di Nanoscienze - CNR, Via Campi 213/A, 41125 Modena, Italy}
\author{S.~Heinze}
\affiliation{Instit\"ut f\"ur Theoretische Physik und Astrophysik, Christian-Albrecht-Universit\"at zu Kiel, Leibnizstr. 15, 24098 Kiel, Germany}
\author{V.~Bellini}
\email[Electronic address: ]{valerio.bellini@ism.cnr.it}
\affiliation{S3 - Istituto di Nanoscienze - CNR, Via Campi 213/A, 41125 Modena, Italy}
\affiliation{Istituto di Struttura della Materia (ISM) - Consiglio Nazionale delle Ricerche (CNR), I-34149, Trieste, Italy}

\begin{abstract}
Using first-principles calculations we demonstrate sizable exchange coupling between a magnetic molecule and a magnetic substrate via a graphene layer. As a model system we consider cobaltocene (CoCp$_2$) adsorbed on graphene deposited on Ni(111).
We find that the magnetic coupling between the molecule and the substrate is antiferromagnetic and varies considerably
depending on the molecule structure, the adsorption geometry, and the stacking of graphene on Ni(111).
We show how this coupling can be tuned by intercalating a magnetic monolayer, e.g. Fe or Co, between graphene and Ni(111). We identify
the leading mechanism responsible for the coupling to be the spatial and energy matching of the frontier orbitals of \cocp\ and graphene close to the Fermi level, and we demonstrate the role of graphene as an electronic decoupling layer, yet allowing spin communication between molecule and substrate.

\end{abstract}
 \maketitle

The emerging field of organic spintronics capitalizes on the novel functionalities achieved when organic molecules are adsorbed on magnetic substrates. The ability to manipulate and tune these functionalities is an important goal. Several problems remain however, before such systems can be incorporated into new technological devices. One in particular is the capability to adsorb molecules on surfaces without any detrimental effects being caused to either the structural or magnetic properties of the molecule.
For this reason it is vital to choose molecules with maximum structural robustness upon adsorption \cite{Mannini2009,Ghirri2011,Kahle2012}. To this end, the phthalocyanine and porphyrin families are popular choices due to their planar geometry \cite{Wende2007,Javaid2010,Waeckerlin2010,Lodirizzini2011,Schwoebel2012,Annese2013}. 
However, in some cases, the strong interaction between the metal ion of such flat molecules and the substrate can modify its electronic states and even quench its magnetic moment \cite{Brede2010}.

The use of non-planer molecules, such as metallocenes, can minimize this effect. Metallocenes are composed of a 3$d$ transition-metal ion sandwiched between two cyclopentadienyls (Cp). Depending on the metal ion species, both non-magnetic and paramagnetic behavior can be found \cite{Xu2003}.
The spin of the metal ion is shielded from the surface by the cage formed by the two Cp rings, reducing the possibility that it will be modified substantially after adsorption. Unfortunately, the deposition of metallocenes on metallic surfaces is a difficult process \cite{Heinrich2011} and, in some cases, complete dissociation of the molecule occurs \cite{Braun2006,Choi2006}.

The intercalation of a graphene spacer layer between the reactive surface and the metallocene can reduce the possibility of molecular dissociation during deposition.
Additionally, evidence of charge transfer at molecule-graphene-Ni(111) interfaces \cite{Dou2011,Uihlein2013} and the theoretical prediction of large charge transfer from cobaltocene  (CoCp$_2$) to graphene  \cite{Li2011} would suggest that a magnetic coupling between cobaltocene and the Ni(111) surface through the graphene layer is still achievable.

In this Letter, we predict, by first principles electronic structure methods, a sizable magnetic coupling for CoCp$_2$ adsorbed on a graphene layer deposited on a Ni(111) substrate.
Furthermore, we propose intercalation of different ferromagnetic metal monolayers, such as Fe and Co, between graphene and the Ni substrate as a route to tailor the magnetic coupling. Due to the unique electronic properties of graphene \cite{Candini2011,Avdoshenko2011}, metal-organic systems of this kind could serve as a basis for future spintronics devices.

Density functional theory (DFT) calculations have been performed using the projector augmented wave method as implemented in the VASP code \cite{kresse1996b,kresse1999} with the Perdew-Burke-Ernzerhof (PBE) exchange correlation functional \cite{perdew1996}. Dispersion interactions have been included according to the DFT-D2 approach  \cite{Grimme2006}. Further computational details can be found in the Supplemental Material \cite{SM}.
\begin{figure}
 \begin{center}
  \includegraphics[width=7cm]{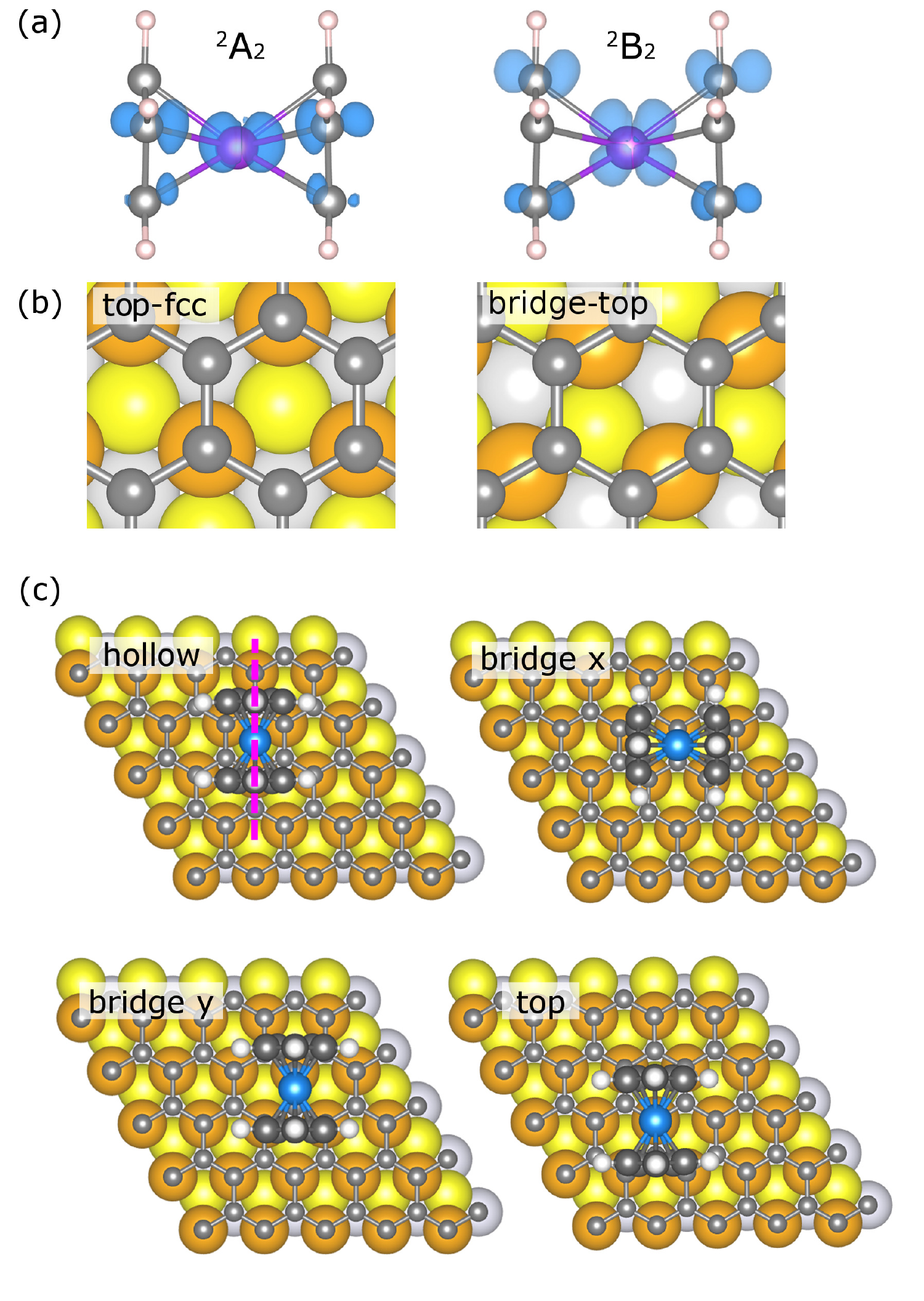}
   \caption{\label{fig:figure_1} (a) Probability density of the CoCp${}_2$ HOMO
   for the   ${}^2$A${}_2$ and ${}^2$B${}_2$ states.
   (b) top-fcc and bridge-top stacking of graphene on Ni(111) (the topmost, second, and third Ni layers are colored orange, yellow, and grey, respectively).
   (c) Adsorption geometries of the CoCp${}_{2}$ on graphene/Ni(111) for the top-fcc stacking (Co, C, H atoms in \cocp\ are colored blue, dark grey, white, respectively).}
  \end{center}
\end{figure}
For a comprehensive characterization of the interface geometry we consider three possible structural degrees of freedom, namely the molecular conformation, the graphene-substrate stacking, and the molecular adsorption site.
Isolated CoCp$_2$ has already been studied extensively by DFT \cite{Xu2003} and several possible structures have been studied.
We consider here \cocp\ in the D${}_{5h}$ high-symmetry configuration \cite{SM} where two possible Jahn-Teller distorted structures characterized by two different electronic states occur.
The probability density of the highest occupied molecular orbital (HOMO) of both of these states, labeled ${}^2$B${}_2$ and ${}^2$A${}_2$, are plotted in Fig. \ref{fig:figure_1}(a). 
In both cases, the \cocp\ molecule attains a nominal S=1/2 spin.
The small lattice mismatch (1.2\%) of graphene and Ni(111) lattice constant results in pseudomorphic growth and the flat conformation of the graphene layer \cite{Dzemiantsova2011}.
DFT calculations have shown that the bonding between graphene and the Ni(111) surface is primarily due to van der Waals (vdW) interactions \cite{Vanin2010}, with a binding distance of $\sim$2.1~{\AA}, in good agreement with experiments \cite{Gamo1997}.
The morphology of the graphene/Ni(111) interface has been investigated experimentally \cite{Zhao2011} and two stable configurations observed (see Fig. \ref{fig:figure_1}(b)). The top-fcc stacking has two inequivalent C atoms, one on top of the Ni(111) surface atom (C${}_{top}$), the other on the fcc site (C${}_{fcc}$). The carbon atoms of the bridge-top stacking are in bridge positions with respect to underlying Ni atoms. We find the top-fcc stacking to be more stable by 6.5 \milli\electronvolt\ per C atom than the bridge-top stacking.
\begin{table}[tb]
\begin{center}
\caption{\label{tab1}
Total energy difference $\Delta E$ (\milli\electronvolt), Co-graphene distance $d$ ({\AA}), and exchange energy $E_{ex}$ (\milli\electronvolt) for different structural and electronic configurations of CoCp${}_{2}$ on graphene/Ni(111) for antiparallel alignment of Co and Ni magnetic moments.}
 \begin{ruledtabular}
\begin{tabular}{lcrcc}
& Config. & $\Delta E$ & $d$ & $E_{ex}$ \\
\hline
\hline
${}^2$B${}_2$, top-fcc, hollow & 1 & 0.0 &  4.31  & $-9.7$ \\
${}^2$A${}_2$, top-fcc, hollow & 2 & +4.6 &  4.31  &  $-1.3$ \\
${}^2$B${}_2$, top-fcc, bridge x & 3 & +55.2 & 4.30  & $-4.6$ \\
${}^2$B${}_2$, top-fcc, bridge y & 4 & +74.8 & 4.31  &  $-8.1$ \\
${}^2$B${}_2$, bridge-top, hollow & 5 & +105.2 & 4.29  & $-9.2$ \\
${}^2$B${}_2$, top-fcc, top & 6 & +147.9 & 4.40 & $-6.8$ \\
\end{tabular}
\end{ruledtabular}
\end{center}
\end{table}
Finally, we find the configuration with the molecule axis parallel to the graphene layer more stable by 40 meV as compared to the case of perpendicular orientation, consistent with Ref. \onlinecite{Li2011}.

We have taken into account several possible adsorption geometries \cocp\ can assume on graphene/Ni(111), which we labelled as hollow, bridge, and top, depending on the position of the Co atom with respect to the C atoms below (Fig. \ref{fig:figure_1}(c)).
The results are presented in Table \ref{tab1}, including the total energy difference $\Delta E$ (with respect to the ground state), the Co-graphene distance $d$ and the exchange coupling energies $E_{ex}$, defined as  $E_{ex} = E_{AP}-E_{P}$, where $E_{AP}$ ($E_{P}$) is the total energy of the system when the spin moment of the Co atom is antiparallel (parallel) to the one of the Ni slab.
Here, a negative value of the exchange energy indicates that the cobaltocene's spin moment preferentially orients antiparallel to the Ni magnetization.
 The lowest energy configuration is found when the molecule is adsorbed on the hollow site of graphene, which has a top-fcc stacking on the underlying Ni(111) substrate.
The calculated adsorption energy of this configuration is $\sim$0.64 \electronvolt, somewhere between those indicating physisorption and chemisorption.
A comparison of the total energies in Table \ref{tab1} shows that, except for the case of configuration 2, all other configurations are strongly energetically unfavorable.
\begin{figure}
 \begin{center}
  \includegraphics[width=8.4cm]{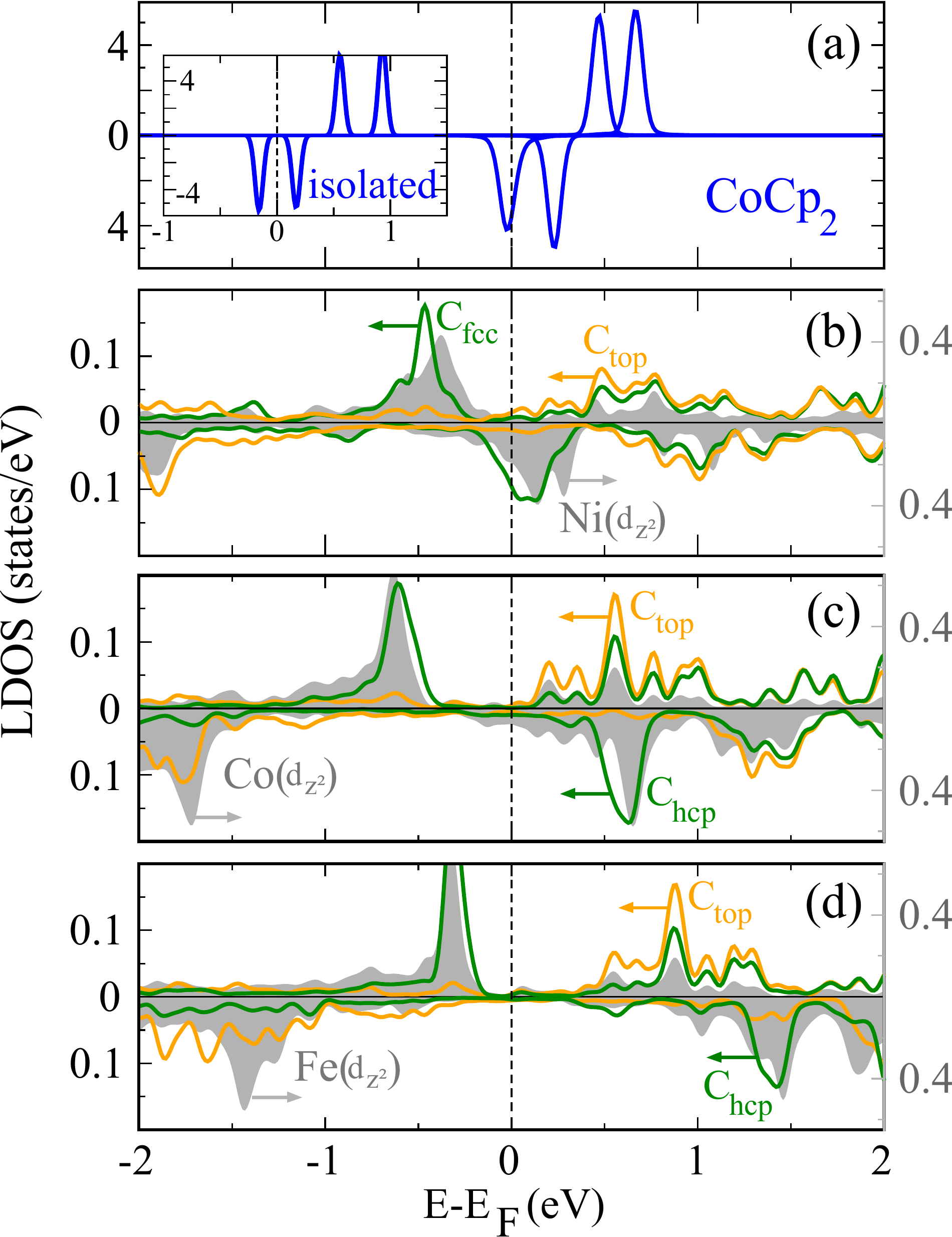}
   \caption{\label{fig:figure_2}
LDOS of CoCp${}_{2}$ on graphene/M/Ni(111) in the antiparallel configuration, with M = Ni, Co and Fe:
(a) 3$d$ states of Co of \cocp\; (b)-(d) 3$d_{z^2}$ states of the M layer atoms and 2$p{}_{z}$ states of C${}_{top}$ and C${}_{fcc}$ (for M=Ni) or
C${}_{top}$ and C${}_{hcp}$ (for M = Co, Fe; see text).
The $d$ states of the M layer are plotted in grey, while graphene C $p$ states are in orange and dark green.
Each curve in panels (b)-(d) is the average over the three atoms of that species closest to the \cocp\ center.
Inset in panel (a):  3$d$ states of Co of \cocp\ for the isolated molecule.
}
  \end{center}
\end{figure}	
For all configurations, the magnetic ground state shows the molecular spin preferentially aligning antiparallel to the Ni magnetisation, with $E_{ex}$ of the order of $-10$ \milli\electronvolt.
This energy is remarkably large if we consider that the distance between the Co and Ni atoms is approximately 6.4~{\AA}. As a comparison, an exchange energy of only 50\,\milli\electronvolt \ was found for chemisorbed Fe porphyrin on Co(100) \cite{Wende2007}, despite the much smaller Fe -- Co distance of 3.5~{\AA}.
Moreover, the values of $E_{ex}$ that we find are high enough to ensure the stability of the spin moments against temperature-induced fluctuations under typical experimental conditions.
In all cases, the relaxed adsorption distance between the Co ion and graphene lies between 4.3 and 4.4~{\AA} and, therefore,
cannot play a strong role in the differing exchange energies.

To elucidate the physical origin of the molecule-substrate exchange coupling we modify independently three possible contributions: the CoCp${}_2$ electronic state, the graphene stacking, and the CoCp${}_2$ adsorption site.
For the first we found that switching from the ${}^2$B${}_2$ to the ${}^2$A${}_2$ electronic configurations (configurations 1 and 2 in Table \ref{tab1}) lowers the exchange energy to $-1.3$ \milli\electronvolt. This considerable decrease can be attributed to the reduced extent of the CoCp${}_2$ spin-polarized HOMO (see Fig. \ref{fig:figure_1}(a)) which is critical to determining the size of the coupling.
Varying the graphene stacking from top-fcc to bridge-top (configurations 1 and 5) does not influence the magnetic coupling in any appreciable way. This is somewhat surprising since the magnetic moment induced on graphene is approximately one order of magnitude smaller in the bridge-top than in the top-fcc stacking, with values of $+0.002~\mu_B$ and $-0.03$ $/$ $+0.02~\mu_B$, respectively.
We can conclude therefore that the magnetic coupling does not depend on the size of the magnetic moment induced on the graphene atoms.
Finally, varying the adsorption site (configurations 1, 3, 4 and 6) can change $E_{ex}$ by up to a factor of two. However, the coupling remains antiferromagnetic in all cases.

In Fig. \ref{fig:figure_2} we present the local density of states (LDOS) of the system in its ground state (configuration 1).
Upon adsorption on the surface, we observe a small shift to higher energies of the molecular Co $d$ orbitals with the result that the HOMO is pinned to the Fermi level (E$_F$) of the substrate. It also becomes partially depopulated. This is accompanied by a charge transfer of 0.28 e$^-$ from the molecule to the surface and a decrease of the magnetic moment size associated to the Co atom from $+0.74~\mu_B$ to $+0.47~\mu_B$.
A hybridization between the 2$p_z$ orbital of the graphene atoms and the 3${d}_{z^2}$ orbital of the Ni atom is also evident, resulting in the polarization of graphene.
Notably, only the C${}_{fcc}$ atoms exhibits
this strong hybridization with the Ni atoms close to E$_F$.
The energy overlap between the minority states of graphene and the minority $d$ states of CoCp${}_2$ just below $E_F$ is responsible for the stabilisation of the AP alignment. This energy matching is absent for the P alignment, due to the inverted HOMO spin polarization.
We can thus conclude that the spin polarisation of graphene close to $E_F$ determines the sign of the magnetic coupling.
This is further corroborated by the analogous situation occurring for configuration 5, for which both the graphene LDOS around $E_F$ \cite{SM} and the magnetic coupling are similar to the ones of configuration 1.
Such a dependence suggests that if one can modify the induced spin polarization of graphene in this energy window, one can modify the magnetic coupling.
In order to validate this idea, we have performed several additional calculations, intercalating different magnetic monolayers (Fe and Co) between graphene and the Ni(111) substrate. 
Experimentally the intercalation of Ni and Co monolayers between graphene and Ir(111) \cite{Pacile2013,Decker2013} and of Fe between graphene and Ni(111) \cite{Weser2011a} has been successfully achieved. As there are no experimental data for graphene/Co/Ni(111), we have used the same structure as for graphene/Fe/Ni(111).
As discussed in Ref. \onlinecite{Weser2011a} the intercalated Fe atoms are preferentially placed in the fcc hollow sites of the Ni, following the Ni(111) stacking. On this substrate, graphene adsorbs in a top-hcp structure, where the two inequivalent graphene C atoms are placed alternatively above the Fe atoms and the hcp sites (corresponding to the topmost Ni layer) \cite{SM}. 

\begin{table}[t]
\begin{center}
\caption{\label{tab2} Magnetic moments of the two non-equivalent atoms of graphene m$_C^{top}$ ($\mu_B$) and m$_C^{fcc/hcp}$ ($\mu_B$), the interface metal monolayer m$_{M}$ ($\mu_B$), and  the exchange energies ${E}_{ex}$ (\milli\electronvolt) for {CoCp${}_2$} on graphene/M/Ni(111) (M = Ni, Co, Fe).}
\begin{ruledtabular}
\begin{tabular}{lccc}
& gr/Ni/Ni & gr/Co/Ni & gr/Fe/Ni \\
\hline
\hline
m$_C^{top}$ & $-0.02$ & $-0.04$ & $-0.05$ \\
m$_C^{fcc/hcp}$ & $+0.03$ & $+0.04$ & $+0.04$ \\
m$_{M}$ & $+0.47$ & $+1.52$ & $+2.39$ \\
$E_{ex}$ & $-9.7$ & $-2.3$ & $+2.0$ \\
\end{tabular}
\end{ruledtabular}
\end{center}
\end{table}

We present in Fig. \ref{fig:figure_2} (c) and (d) the LDOS in the case of Fe and Co intercalation.
The values of the corresponding magnetic moments and the exchange energies are listed in Table \ref{tab2}.
The magnitude of the spin moment in the interface metal (M) layer increases as one goes from Ni to Co to Fe and, due to hybridization, this increase also occurs for the moments induced on the C atoms, i.e., m$_C^{top}$ and m$_C^{fcc}$.
Counterintuitively, the magnetic coupling is not found to increase in line with the magnetic moment and in fact decreases. We can identify a trend for the exchange energy between \cocp\ and the investigated substrate from large antiferromagnetic ($E_{ex}=-9.7~ \milli\electronvolt$) for graphene/Ni(111), to weak antiferromagnetic ($E_{ex}=-2.3~\milli\electronvolt$) for graphene/Co/Ni(111) and weak ferromagnetic ($E_{ex}=+2.0~\milli\electronvolt$) for graphene/Fe/Ni(111).
The energy matching between the HOMO of CoCp$_2$ and the p$_z$ states of the carbon atoms, which drives the coupling between the molecule and substrate, is disrupted by the intercalation of the metal layer. The minority d$_{z^2}$ states of the Co layer lie at higher energies than those of Ni and the Fe states are found at even higher energies.
Due to hybridisation, the p$_z$ orbitals of the graphene atoms are similarly shifted to higher energies. This reduces (for Co intercalation) and finally prevents (for Fe intercalation) the energy matching of the C states with the spin-polarized HOMO of CoCp$_2$ with a resultant decrease in the magnetic coupling.

\begin{figure}
 \begin{center}
  \includegraphics[width=7cm]{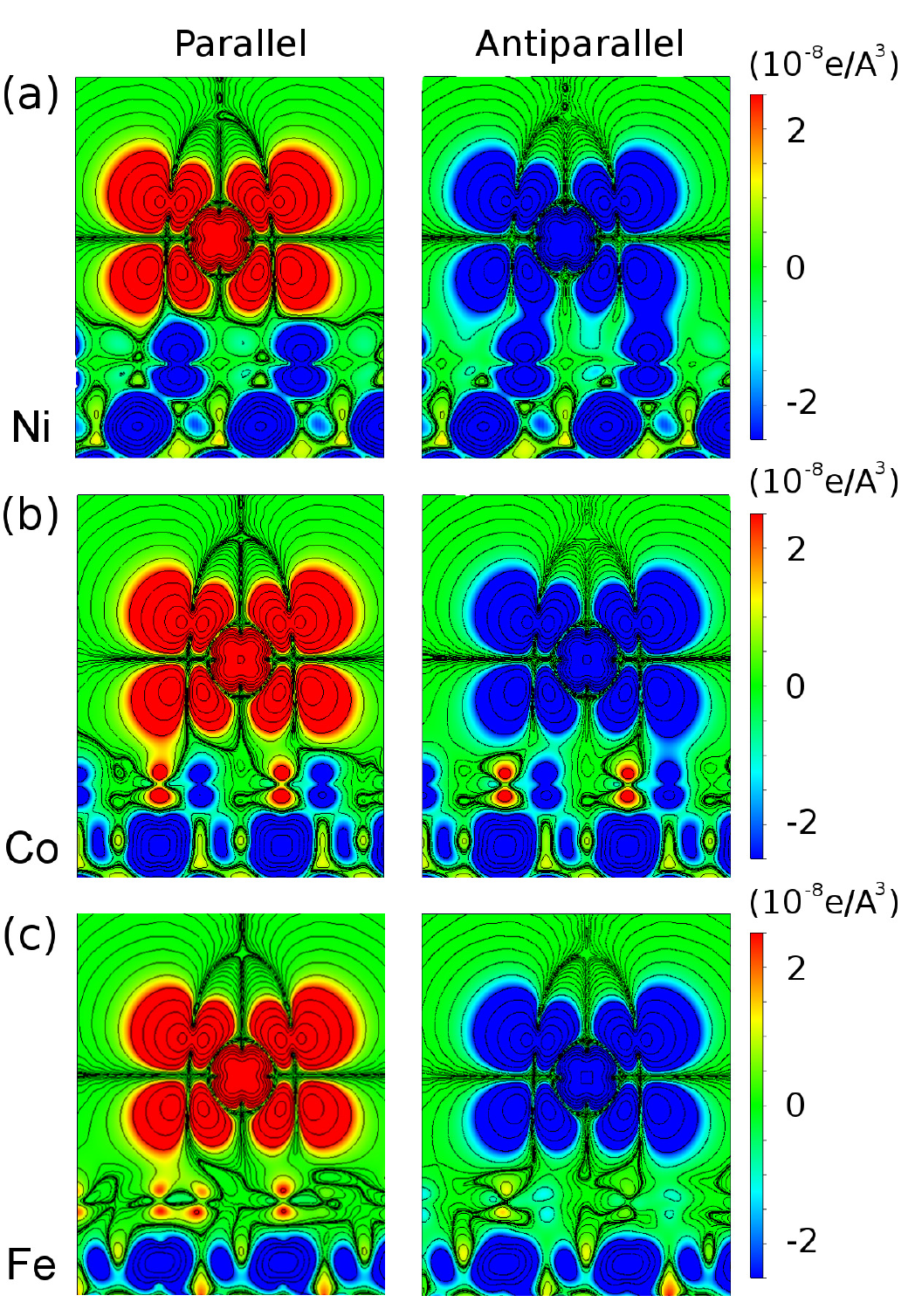}
   \caption{\label{fig:figure_3}
   Cross-sectional plots of the local magnetization density integrated from $-0.1$ \electronvolt\ to the Fermi level of {CoCp${}_2$} on
   (a) graphene/Ni(111), (b) graphene/Co/Ni(111), and (c) graphene/Fe/Ni(111). Left (right) panels refer to the parallel (antiparallel) configuration. 
The cross sectional plane, indicated as purple broken line in Fig. \ref{fig:figure_1}(c) (hollow adsorption site), cuts through the Co atom and is perpendicular to both the substrate and the Cp rings.}
  \end{center}
\end{figure}

Cross sectional plots of the magnetization density, i.e. the difference between the spin up and down charge densities, close to the Fermi level are given in Fig. \ref{fig:figure_3}. 
In panel (a) the spatial matching of the CoCp${}_2$ HOMO with the ${p_z}$ orbitals of the C${}_{fcc}$ atom of graphene adsorbed on Ni(111) is evident in the AP alignment where spin density lobes from the molecule and the surface atoms merge.
In contrast, it is absent in the P alignment resulting in a negative exchange energy.
For CoCp${}_2$ on graphene/Co/Ni (Fig. \ref{fig:figure_3}(b)) there is an excess of majority spin for the C${}_{top}$ atoms and of minority spin for the C${}_{hcp}$ atoms which almost cancel each other. 
However, a small preference towards communication through the minority spins is suggested by the plot in accordance with the weak antiferromagnetic coupling.
For the Fe intercalated layer (Fig. \ref{fig:figure_3}(c)), the spin density in the graphene indicates spin communication for the P alignment, but not for the AP, explaining the positive exchange energy.
The analysis performed for configuration 5 of Tab. \ref{tab1} (see Fig. 4 in Ref. \onlinecite{SM}) reveals a scenario similar to the one of Fig. \ref{fig:figure_3}(a) and is consistent with an exchange coupling of similar size. In contrast in configuration 2 spatial overlap between \cocp\ and graphene states at E$_F$ is absent, which explains the much weaker magnetic interaction as compared to configuration 1.

Further clarification of the role of graphene in this system can be found by considering the situation when CoCp${}_{2}$ is adsorbed directly on the Ni(111) surface. For the optimized distance between the Co and Ni atoms ($d$ = 4.3~{\AA}) we find a charge transfer of 0.64 e$^-$ from the molecule to the surface which results in a complete quenching of the molecular spin. 
This can be recovered by rigidly shifting the molecule away from the surface by 1~{\AA} whereby the Co ion attains a magnetic moment of $+0.26~\mu_B$. As for the case including the graphene layer, the magnetic coupling is antiparallel, albeit weakly ($E_{ex}=-0.4~\milli\electronvolt$). 
A further rigid shift of the molecule by 1~{\AA} results in an increase of the Co magnetic moment to $+0.35~\mu_B$ while the exchange coupling becomes negligibly small. 

In conclusion, our work demonstrates that graphene plays a vital role in determining the interaction between a magnetic molecule and a ferromagnetic substrate,  behaving as an electronic decoupling layer, yet allowing spin communication.

This work has been supported by FCRM ``THE-SIMS'', the Deutsche Forschungsgemeinschaft via the SFB 677, the European Science Foundation (ESF) under the EUROCORES Program Euro-GRAPHENE, and the German Academic Exchange Service (DAAD).
We acknowledge the CINECA and HLRN centers for granting the high-performance computing resources.

\bibliographystyle{apsrev}

\end{document}


\title{SUPPLEMENTARY MATERIAL \\[1cm] Graphene-mediated exchange coupling between cobaltocene and magnetic substrates}
\date{\today}

\author{S.~Marocchi}
\affiliation{Dipartimento di Fisica, Universit\'a di Modena e Reggio Emilia, Via Campi 213/A, 41125 Modena, Italy.}
\affiliation{S3 - Istituto di Nanoscienze - CNR, Via Campi 213/A, 41125 Modena, Italy}
\author{P.~Ferriani}
\affiliation{Instit\"ut f\"ur Theoretische Physik und Astrophysik, Christian-Albrecht-Universit\"at zu Kiel, Leibnizstr. 15, 24098 Kiel, Germany}
\author{N.~M.~Caffrey}
\affiliation{Instit\"ut f\"ur Theoretische Physik und Astrophysik, Christian-Albrecht-Universit\"at zu Kiel, Leibnizstr. 15, 24098 Kiel, Germany}
\author{F.~Manghi}
\affiliation{Dipartimento di Fisica, Universit\'a di Modena e Reggio Emilia, Via Campi 213/A, 41125 Modena, Italy.}
\affiliation{S3 - Istituto di Nanoscienze - CNR, Via Campi 213/A, 41125 Modena, Italy}
\author{S.~Heinze}
\affiliation{Instit\"ut f\"ur Theoretische Physik und Astrophysik, Christian-Albrecht-Universit\"at zu Kiel, Leibnizstr. 15, 24098 Kiel, Germany}
\author{V.~Bellini}
\affiliation{S3 - Istituto di Nanoscienze - CNR, Via Campi 213/A, 41125 Modena, Italy}
\affiliation{Istituto di Struttura della Materia (ISM) - CNR, c/o Sincrotrone Trieste SCpA,
S.S. 14 Km 163.5, I-34149, Trieste, Italy}

 \maketitle

\begin{center}
\textbf{\large{Computational details}}
\end{center}

We performed preliminary calculations on a 
p(1 $\times$ 1) hexagonal supercell to obtain geometry optimized graphene/M/Ni(111) (M = Ni, Co, Fe) structures and a density of states analysis of the systems. In this first part of our work we used a system containing one layer (two C atoms) of graphene and a four-layer metal slab.
An optimized $\Gamma$-centered \textit{k}-point grid of 17 $\times$ 17 $\times$ 1 has been used for these small supercell calculations. Graphene layer and the top layer of metal were relaxed along the z-axis. 
The three bottom layers of Ni were  
fixed at the bulk geometry with a lattice parameter of 2.49~{\AA} that corresponds to the PBE bulk optimized Ni-Ni distance. The geometries of these small  cells have been replicated in plane in order to build the bigger p(5 $\times$ 5) supercell containing not only the graphene-metal slab but also the cobaltocene (CoCp${}_{2}$) molecule. In these second set of calculations, the graphene and metal layers were  
fixed and only the CoCp${}_{2}$ was fully relaxed. A $\Gamma$-centered grid of 3 $\times$ 3 $\times$ 1 k-points has been used. Minimization proceeded till forces were lower than 0.01 \electronvolt\ {\AA}$^{-1}$. 

We included in our calculations the van der Waals (vdW) dispersion interactions. It is known that the non-local exchange-correlation energy functional (vdW-DF) results in an equilibrium distance between graphene and the topmost Ni layer larger than 3.5~{\AA} \cite{Vanin2010} for all the calculated structures, in disagreement with experiments. Thus we employed the semi-empirical potential DFT-D2 of Grimme \cite{Grimme2006}.
With this method the estimate of the interaction energy is less reliable as compared to the vdW-DF, however the typical errors never exceed 20 \% \cite{Grimme2006}. On the other hand, the DFT-D2 gives trustworthy equilibrium distances. On average the distances obtained with this method for small aromatic systems underestimate by about 5 \% the experimental values \cite{Grimme2006}. In light of the fact that in our calculations it is essential to get the equilibrium distances as reliable as possible, we decided to use the semi-empirical potential DFT-D2. 

\

\begin{center}
\textbf{\large{Cobaltocene electronic structure and adsorption on graphene/Ni(111)}}
\end{center}

\begin{figure}[htbp]
 \begin{center}
  \includegraphics[width=0.5\linewidth]{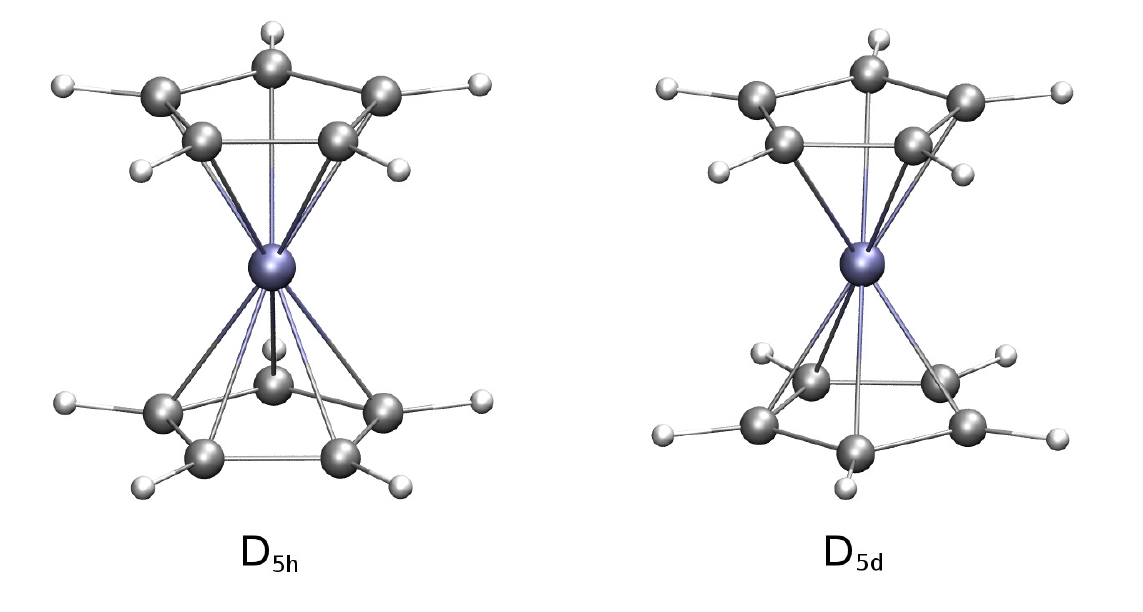}
   \caption{\label{fig:struc_D5h_D5d}}
  \end{center}
\end{figure}

We plot in Fig. \ref{fig:struc_D5h_D5d} the two possible structures of a metallocene molecule,
D$_{5h}$ and D$_{5d}$, named after their point group symmetry. In the former  the two pentagons of the Cp rings
are symmetric, 
while in the latter they are rotated by 180$^\circ$.
Total energy calculations show that D${}_{5h}$ symmetry is energetically favored for isolated CoCp${}_2$ molecules, with respect
to D${}_{5d}$.
The crystal field produced by the Cp rings split the 3$d$
orbitals of Co. If we consider the axis of the molecule as our z-axis,
the seven 3$d$ electrons are split in: (i) two
electrons in a singlet orbital derived from d${}_{z^2}$, (ii) four
electrons in a doubly degenerate orbitals derived from d${}_{xy}$ and d$_{x^2-y^2}$, (iii) one electron in a doubly degenerate orbitals derived
from d${}_{yz}$ and d${}_{xz}$. The D${}_{5h}$ symmetry is distorted to
remove the degeneracy of the frontier orbitals, according to the
Jahn-Teller effect, so the symmetry is reduced to C${}_{2v}$. There are
two possible distortions and thus two possible electronic states;
depending on whether the Cp rings tilt slightly toward the molecular
center or outwards, the ${}^2$B${}_2$ or the ${}^2$A${}_2$ electronic
states are respectively produced, as shown in Fig. 1 panel (a) in the main text.
In the ${}^2$B${}_2$ state, the HOMO has d${}_{yz}$ character while in the ${}^2$A${}_2$ state it has
d${}_{xz}$ character. 

The adsorption of CoCp${}_2$ on graphene/Ni(111) can be regarded as strong physisorption as inferred from the fact that (i) the smallest distance between the H atoms of CoCp${}_2$ and the C atoms of graphene is $\sim$2.4~{\AA}, (ii) there is no appreciable distortion of the structure of CoCp${}_2$ upon deposition on the surface, and (iii) the C -- C, C -- Ni and C -- H bonds are non-polar (or only weakly polar) excluding the formation of hydrogen bonds.

The configuration where the CoCp${}_2$ is oriented with its axis perpendicular to the surface and in the top adsorption site, i.e. where the Co atom is directly above one of the graphene C${}_{top}$ atom, 
is less stable than the configuration with the CoCp${}_{2}$ axis parallel to the substrate surface (configuration 1 in Tab. 1 in the main text) by 40 \milli\electronvolt \ . The minimum distance is 3.0~{\AA}, as compared to 2.4~{\AA}.

\
\begin{center}
\textbf{\large{Graphene stacking for Fe and Co intercalation}}
\end{center}

\begin{figure}[htbp]
 \begin{center}
 \includegraphics[width=0.5\linewidth]{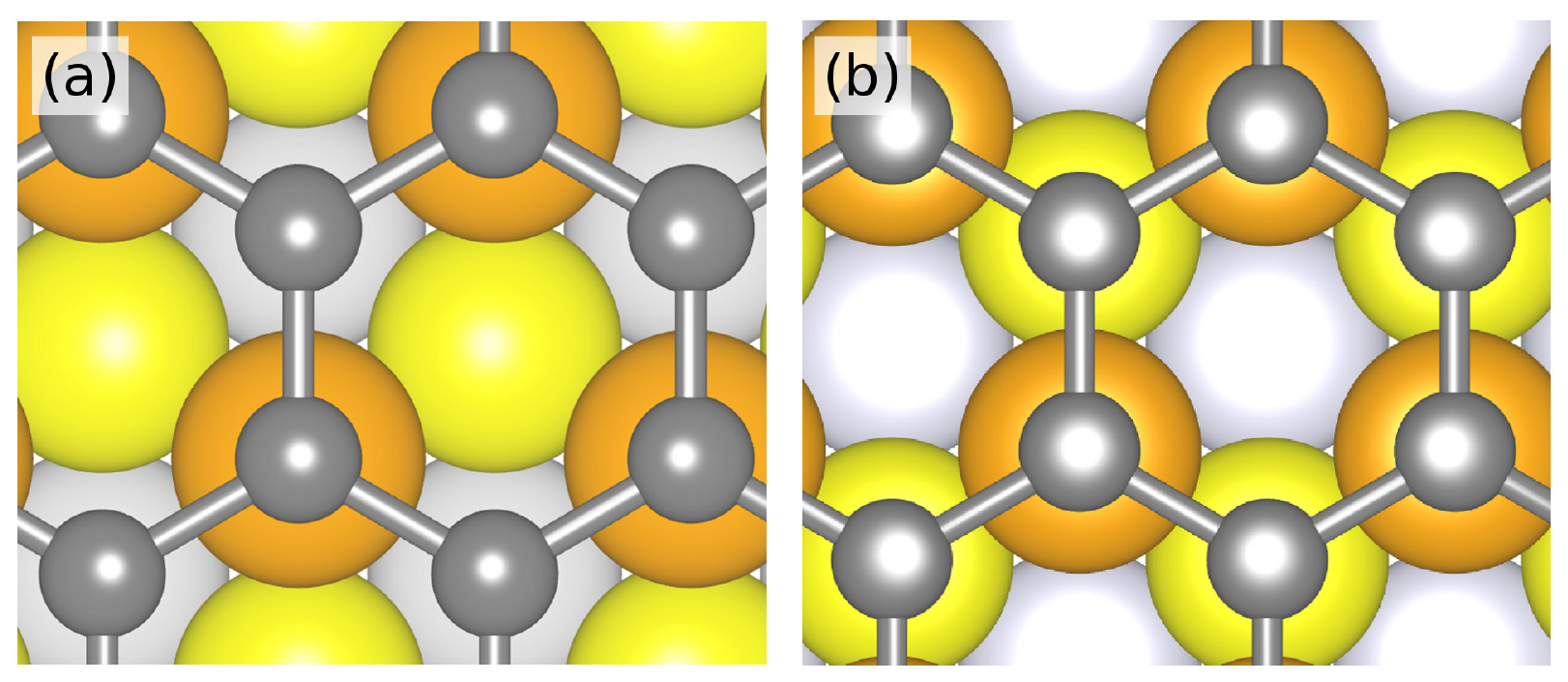}
   \caption{\label{fig:struc_top-hcp}}
  \end{center}
\end{figure}

The surface lattice constant of the Ni(111) is 2.49~{\AA}, with a very small lattice mismatch, i.e. 1.2 \%, with the one of graphene ( 2.46 ~{\AA} ). In the case of the intercalated Fe or Co monolayer, lattice mismatch increases, yet a single layer is stable and does not lead
to relevant modification in the graphene structure. 
The energetically favored top-fcc structure for graphene/Ni(111) (a) compared with the top-hcp structure (b) attained for graphene/M/Ni(111), with M = Fe and Co; first (topmost, orange), second (yellow), and third (light gray) metallic layer atoms below graphene atoms (dark grey) are depicted in Fig. \ref{fig:struc_top-hcp}. This figure has been obtained using VESTA \cite{Mommako5060}

\
\begin{center}
\textbf{\large{Analysis of the LDOS and cross sectional plots close to the Fermi level}}
\end{center}

\begin{figure}[htbp]
 \begin{center}
  \includegraphics[width=0.7\linewidth]{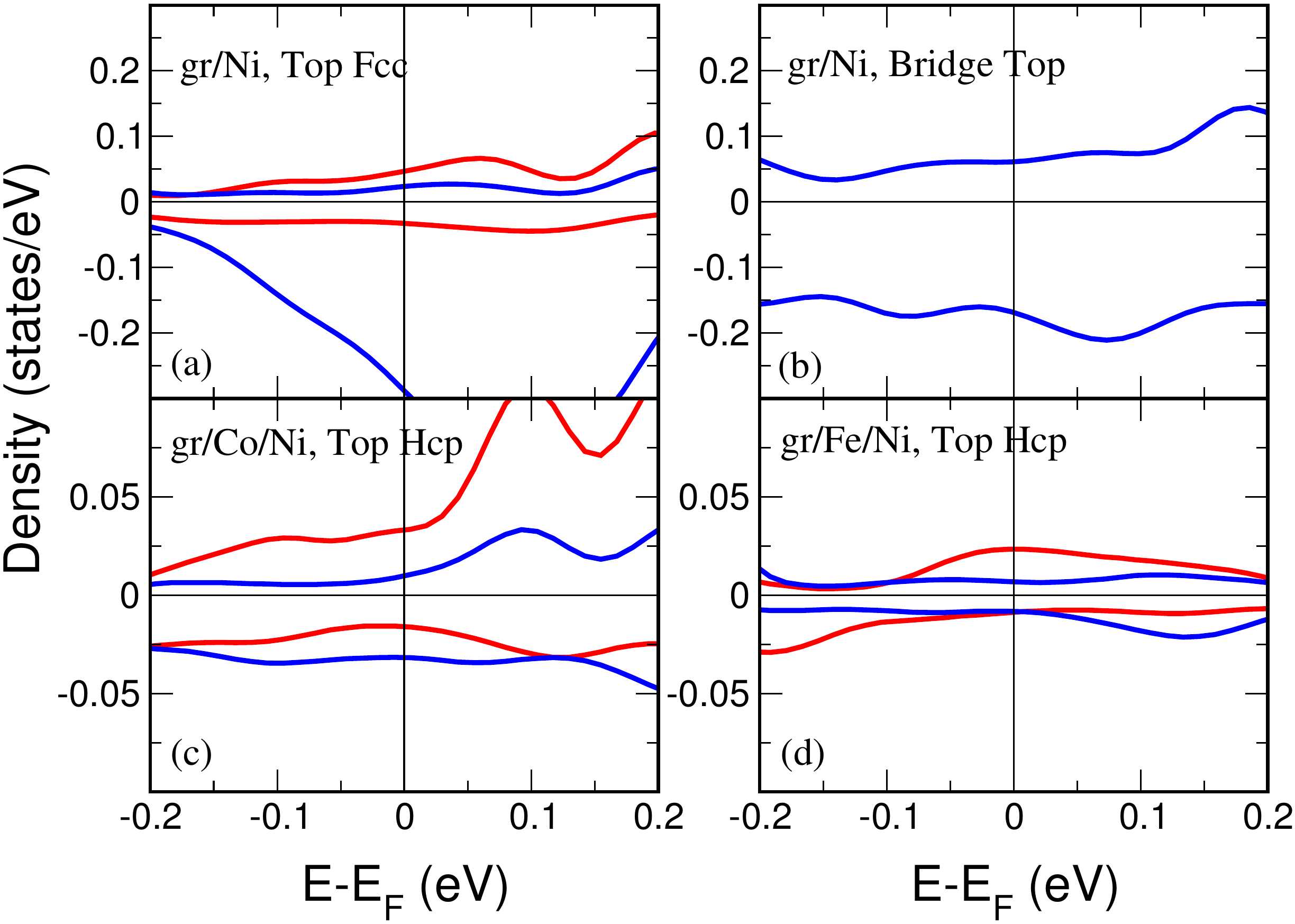}
   \caption{\label{fig:doszoom}}
  \end{center}
\end{figure}

We plot in Fig. \ref{fig:doszoom} the spin-polarized LDOS at the C graphene atoms, focussing on 
the region close to the Fermi level (E$_F$), for configuration 1 and 5 of Tab. 1 in the main text,
and the systems with Co and Fe intercalation.
The two curves in panel (a), (c) and (d) represent the two inequivalent C atoms (C${}_{top}$ in red, C${}_{fcc}$ and C${}_{hcp}$ in blue);
in the configuration of panel (b), only one inequivalent C atom is present. Looking at panels (a) and (b), it is
evident that close to E$_F$ spin polarization of the LDOS
is similar for top-fcc and bridge-top, although magnetic moments induced on C atoms are one order of magnitude apart, i.e. $+0.020$ {\it vs.} $+0.002~\mu_B$. This leads to a similar (in size) antiparallel magnetic coupling between the molecule and the substrate spins. In panel (c), Co intercalation leads to a reduction of the spin polarization; yet the area subtended in the region within 0.1 \electronvolt\ below E$_F$ in the minority spin channel ($\sigma=\downarrow$) is larger than the one in the majority channel ($\sigma=\uparrow$), and the matching with the CoCp$_2$ HOMO results still in a (smaller) antiparallel coupling. For Fe intercalation, panel (d), in the same energy region, there is a small excess of majority spin, which is consistent with the small parallel coupling between the molecule and the substrate spins.

\begin{figure}[htbp]
 \begin{center}
  \includegraphics[width=0.7\linewidth]{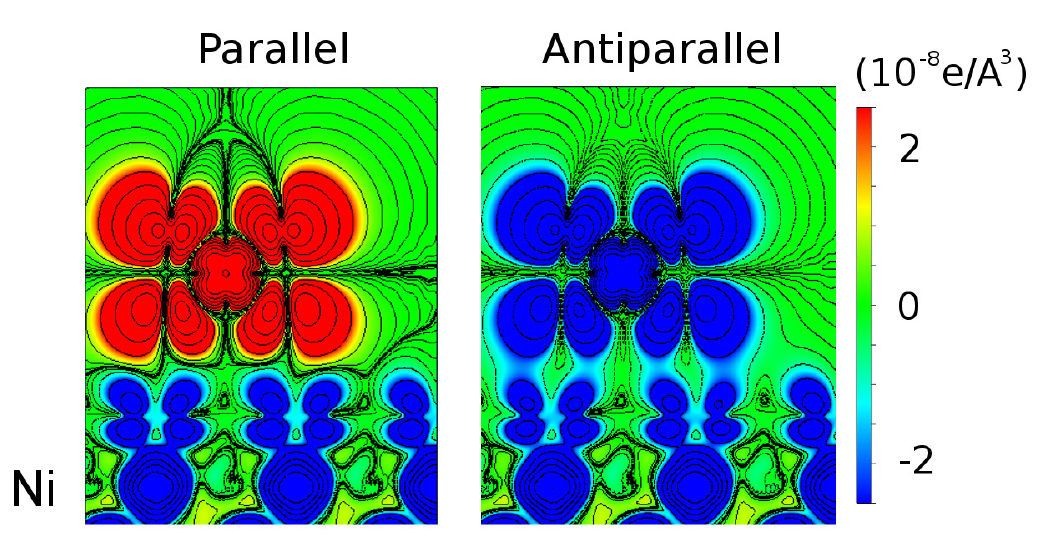}
   \caption{\label{fig:bridge-top_cp}}
  \end{center}
\end{figure}

The Fig. \ref{fig:bridge-top_cp} shows the cross-sectional plot of the local magnetization density (difference between the spin up and down charge densities) integrated from $-0.1$ \electronvolt\ to E$_F$ of {CoCp${}_2$} on graphene/Ni(111) for the bridge-top stacking. The presence of spin density between the graphene layer and the cobaltocene indicates spin communication for the case of antiparallel alignment of the molecule and substrate magnetization. This is missing in the parallel case consistent with the negative spin polarization of the LDOS showed in Fig. \ref{fig:doszoom}(b). This situation is analogous to the one of the top-fcc stacking discussed in the main text and explains the similar exchange energy (see Tab. 1 in the main text).

